\documentclass[conference,10pt,draftcls,onecolumn]{IEEEtran}

\usepackage[final]{graphicx}
\usepackage{amsmath,amsfonts,amssymb,cite,setspace,color,bm}
\usepackage[lmargin=0.65in,rmargin=0.65in,bmargin=1.1in,tmargin=0.75in]{geometry}
\usepackage{units}

\allowdisplaybreaks
\graphicspath{{fig/}}

\def\nn{\nonumber}

\newtheorem{mytheo}{Theorem}

\begin{document}
\title{Degrees-of-Freedom Regions for $K$-User MISO Time-Correlated Broadcast Channel}
\author{Yi Luo$^{*}$, Tharmalingam Ratnarajah$^{*}$ and Anastasios K. Papazafeiropoulos$^{\dag}$\\
$^{*}$Institute for Digital Communications, The University of Edinburgh, Edinburgh U.K.\\
$^{\dag}$Communications and Signal Processing Group, Imperial College London, London, U.K.\\
Email: \{y.luo; t.ratnarajah\}@ed.ac.uk, a.papazafeiropoulos@imperial.ac.uk
}

\maketitle
\begin{abstract}
In this paper, we study the achievable degrees-of-freedom (DoF) regions of the $K$-user multiple-input-single-output (MISO) time correlated broadcast channel (BC). The time correlation induces knowledge of the current  channel state information at transmitter (CSIT) with an estimation error $P^{-\alpha}$, where $P$ is the signal-to-noise ratio (SNR). We consider the following two scenarios: $(i)$ $K$-user with $K$-antenna base station (BS) and $(ii)$ $3$-user with $2$-antenna BS. In case of symmetric DoF tuples, where all the users obtain the same DoF, we derive the total DoF equal to $\frac{K(1-\alpha)}{1+\frac{1}{2}+\cdots+\frac{1}{K}}+K\alpha$ for the first scenario and $\frac{3+\alpha}{2}$ for the second one. In particular, we provide the achievability schemes for these two DoF tuples. Nevertheless, we also consider the asymmetric case where one of the users is guaranteed {\it one} DoF, and provide the achievability scheme. Notably, the consistency of the proposed DoF regions with an already published outer bound , as well as with the Maddah-Ali-Tse (MAT), which assumes only perfect delayed CSIT, and the ZF beamforming schemes  (perfect current CSIT) consents to the optimality of the proposed achievability schemes.
\end{abstract}

\section{Introduction}  \label{sec:introduction}
As interference is a critical factor for the performance of modern wireless communication, its management by using practically available (imperfect) channel state information at transmitter (CSIT) is becoming an important topic. There are many interference management techniques that have been proposed in the literature, including interference alignment (IA) \cite{Cadambe2008} and beamforming (e.g., zero-forcing (ZF)). Most of the beamforming techniques like IA and ZF can achieve high degrees of freedom (DoF) but require high quality of current CSIT. Unfortunately, the time division multiple access (TDMA) scheme requires no CSIT but can only achieve {\it one} DoF in total.

As high-quality current CSIT is difficult to achieve by  using modern CSI feedback techniques \cite{Caire2010}, the trade-off between CSIT requirement and achievable DoF becomes an interesting topic. The MAT scheme proposed in \cite{Maddah2012} exploits completely delayed CSIT and achieves higher DoF than the schemes in absence of any CSIT, i.e., the completely delayed CSIT is useful to achieve higher DoF. Recently, given that knowledge of only delayed CSIT is pessimistic, information-theoretic studies have successfully extended the idea of the MAT scheme to a  time-correlated channel, where the current CSIT can be estimated by taking into account for  the delayed observations. Specifically, in \cite{Sheng2013} the DoF region is studied for a $2$-user multiple-input-single-output (MISO) broadcast channel (BC), where imperfect current and perfect delayed CSIT are available. Moreover, the same authors extended the MISO BC to the  $2$-user multiple-input-multiple-output (MIMO) interference channel (IC), by considering different antenna settings~\cite{Yi2013_2}. Furthermore, the DoF region is obtained in \cite{luo2014degrees,Chen2013_3} for a $2$-user MISO channel, when both imperfect current and delayed CSIT are known. It is worthwhile to mention that most of all the  previous works, based on the assuption of a time-correlated channel, have  considered only $2$ users. Similarly to the MAT scheme, which has considered a scenario with more than $2$ users, it is interesting to investigate the achievable DoF regions in time-correlated channels with multiple users. Although in \cite{De2013} the DoF outer bound was proposed for a $K$-user and $K$-antenna BS scenario, they were unable to show the achievability scheme reaching this bound.

In this paper, we consider a MISO time-correlated BC,  where CSI is conveyed back to the BS by using a one-time-slot delay link. In this way, imperfect current CSIT can be predicted by using the obtained delayed samples. The estimation error of the current CSIT is $P^{-\alpha}$, where $P$ is the signal-to-noise ratio (SNR). Two scenarios are addressed  in this paper and together with \cite{Sheng2013}, all the available schemes given in \cite{Maddah2012} are successfully extended to a time-correlated channel. Specifically, the cases of $K$-user with $K$-antenna base station (BS) and $3$-user with $2$-antenna BS are studied. In the former setting, $\frac{K(1-\alpha)}{1+\frac{1}{2}+\cdots+\frac{1}{K}}+K\alpha$ DoF are achieved and evenly allocated to $K$ users. Meanwhile, for the asymmetric scheme where one user is guaranteed {\it one} DoF, all the other users achieve $\alpha$ DoF. In the latter setting, $\frac{3+\alpha}{2}$ DoF are  achieved and evenly allocated to $3$ users. It is worth mentioning that all the DoF regions are consistent with  well-known results. That is to say, when the quality of current CSIT is low, the DoF value approaches the DoF achieved by the MAT scheme. When the quality of the current CSIT is high, the achievable DoF value approaches to {\it one} per user, obtained by means of ZF beamforming. Most importantly, we show that our proposed achievability schemes reach the already known DoF outer bounds.

The remainder of this paper is organized as follows. Section \ref{sec:case1} starts with the study of the symmetric case of a $3$ users and $3$-antenna BS, and then the  generalization to the $K$-user and $K$-antenna BS scenario is presented. Interestingly, Section \ref{sec:case2} obtains the DoF for a special assymetric case, i.e., a $3$-user and $2$-antenna BS transmission scenario. Finally, the conclusions are given in Section \ref{sec:conclusion}.

\section{System Model}  \label{sec:model}
We consider a $K$-user MISO BC, where the BS is equipped with $N$ antennas. For Rx-$i$, the received signal at time slot $t$ is given by
\begin{equation}
  y_i(t) = \mathbf h_i(t)\mathbf x(t) + n_i(t),
\end{equation}
where $\mathbf h_i(t)\in\mathbb C^{1\times N}$ denotes the channel vector from  the BS to Rx-$i$, $\mathbf x(t)\in\mathbb C^{N\times 1}$ is the beamformed symbol vector at the BS and $n_i(t)$ is the zero-mean unit-variance additive white Gaussian noise at Rx-$i$.

We assume that the BS obtains perfect CSI with a one-time-slot delay link. Based on the delayed CSIT, the BS will predict the current CSIT to Rx-$i$ at time slot $t$ as $\hat{\mathbf h}_i(t)$. The relation between the estimated current CSIT and the actual one is
\begin{equation}
  \mathbf h_i(t) = \hat{\mathbf h}_i(t) + \tilde{\mathbf h}_i(t),
\end{equation}
where the estimation error vector $\tilde{\mathbf h}_i(t)$ consists of independent identically distributed (i.i.d) Gaussian entries with {\it zero} mean and $P^{-\alpha}$ variance. Note that $P$ is the SNR and parameter $\alpha$ represents the quality of the estimated channel at high SNR. Specifically, $\alpha\!=\!0$ indicates no current CSIT, while $\alpha \!\to\! \infty$ denotes perfect instantaneous CSIT. However, in the case that $\alpha\!\geq\!1$, zero-forcing (ZF) beamforming may achieve no DoF loss due to sufficient knowledge of imperfect current CSIT. Thus, only $\alpha \!\in\! [0,1]$ will be considered.

\section{Symmetric Case $\left( N=K \right)$}  \label{sec:case1}
In this section, DoF regions as well as the achievability schemes for the $3$-user and $3$-antenna BS case is studied first and then extended to the $K$-user and $K$-antenna BS scenario.

\subsection{DoF region for $K=N=3$}
\begin{mytheo}
The achievable DoF region of a $3$-user MISO BC with $3$-antenna BS, using imperfect current CSIT as well as perfect delayed CSIT, is given by Fig.~\ref{fig:region3d_1}.
\begin{figure}[t]
  \centering
  \includegraphics[width=0.9\textwidth]{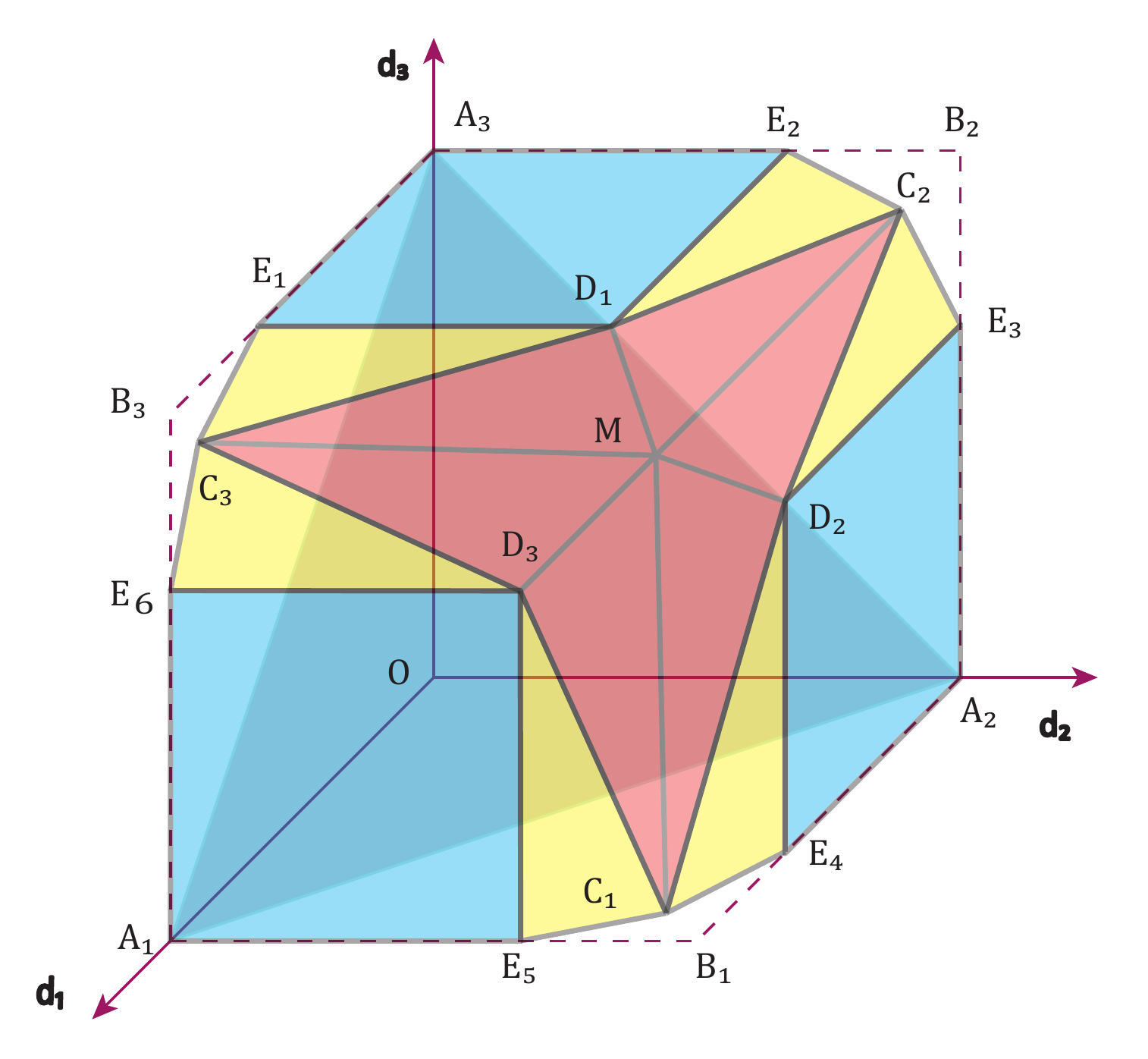}\\
  \vspace{-4mm}
  \caption{The DoF polygon for the symmetric case  $\left( N=K=3 \right)$.}
  \vspace{-4mm}
  \label{fig:region3d_1}
\end{figure}
The coordinates of important vertices are listed below:
\begin{align}
  &M=\left(\frac{6+5\alpha}{11},\frac{6+5\alpha}{11},\frac{6+5\alpha}{11}\right),  \notag \\
  &D_1=(\alpha,\alpha,1), D_2=(\alpha,1\,\alpha),D_3=(1,\alpha,\alpha), \notag \\
  &C_1=\left(\frac{2+\alpha}{3},0,\frac{2+\alpha}{3}\right), C_2=\left(0,\frac{2+\alpha}{3},\frac{2+\alpha}{3}\right), \notag \\
  &C_3=\left(\frac{2+\alpha}{3},0,\frac{2+\alpha}{3}\right). \notag
\end{align}
\end{mytheo}

\textbf{Remark:} The vertices $C_1,C_2,C_3$ are achievable by using the scheme proposed in \cite{Sheng2013}. Vertices $M$, $D_1$, $D_2$, and $D_3$ achieve $3$ DoF when $\alpha=1$, which is achieved by ZF beamforming approach with perfect current CSIT. Vertex $M$ achieves $\frac{18}{11}$ DoF when $\alpha=0$, which are achieved by the MAT scheme. In particular, vertex $M$ can be achieved by using Scheme $\mathcal X_1$, which consists of $3$ phases and totally $11$ time slots. Vertices $D_1,D_2,D_3$ are achieved by a one-time-slot scheme, Scheme $\mathcal X_2$. The achievability schemes will be discussed in detail as follows.

\subsection{Scheme $\mathcal X_1$  (vertex $M$)}
\subsubsection{Phase~$1$}
In the first phase of scheme $\mathcal X_1$, new symbol vectors for Rx-$1$, Rx-$2$ and Rx-$3$ are prepared as $\mathbf s_1$, $\mathbf s_2$, $\mathbf s_3$, respectively, where
\begin{equation}
  \mathbf s_1=\left[
                \begin{array}{c}
                  s_1^{(1)} \\
                  s_1^{(2)} \\
                  s_1^{(3)} \\
                \end{array}
              \right],
  \quad
  \mathbf s_2=\left[
                \begin{array}{c}
                  s_2^{(1)} \\
                  s_2^{(2)} \\
                  s_2^{(3)} \\
                \end{array}
              \right],
  \quad
  \mathbf s_3=\left[
                \begin{array}{c}
                  s_3^{(1)} \\
                  s_3^{(2)} \\
                  s_3^{(3)} \\
                \end{array}
              \right]. \notag
\end{equation}
Each symbol contains $(1-\alpha)\log P$ bits of information, and $\mathbf E[|s_i^{(j)}|^2]\doteq P, i,j=1,2,3$. The symbol $f(x)\doteq g(x)$ means $\lim_{P\to\infty}\frac{f(x)}{g(x)}=\mathcal C$ where $\mathcal C$ is a constant not scaled of $P$. When transmitting the aforementioned symbol vectors, other accompanying symbols are sent, which are denoted as $s_1^{(j)}$, $s_2^{(j)}$, and $s_3^{(j)}$, where $j=4,5,6$. Each of the accompanying symbol contains $\alpha\log P$ bits of information and $\mathbf E[|s_i^{(j)}|^2]\doteq P^\alpha, i=1,2,3,j=4,5,6$. The accompanying symbol $s_i^{(j)}$ is sent at time slot $j-3$. For example, the symbol vectors sent in the first three time slots are given by
\begin{align}
  \mathbf x(1) =&\  s_1 + \mathbf q_{2,3}^\perp(1) s_1^{(4)} + \mathbf q_{1,3}^\perp(1) s_2^{(4)} + \mathbf q_{1,2}^\perp(1) s_3^{(4)}, \notag \\
  \mathbf x(2) =&\  s_2 + \mathbf q_{2,3}^\perp(2) s_1^{(5)} + \mathbf q_{1,3}^\perp(2) s_2^{(5)} + \mathbf q_{1,2}^\perp(2) s_3^{(5)}, \notag \\
  \mathbf x(3) =&\  s_3 + \mathbf q_{2,3}^\perp(3) s_1^{(6)} + \mathbf q_{1,3}^\perp(3) s_2^{(6)} + \mathbf q_{1,2}^\perp(3) s_3^{(6)}, \notag
\end{align}
where $\mathbf q_{i,j}^\perp(t)\in\mathbb C^{3\times 1}$ denotes the normalized beamforming vector lying in the null space of $\mathbf h_i(t)$ and $\mathbf h_j(t)$. At Rx-$i$, the received signal in the first time slot can be expressed as
\begin{align}
  y_i(1) =& \mathbf h_i(1)\mathbf s_1 + \mathbf h_i(1)\mathbf q_{2,3}^\perp(1)s_1^{(4)}+ \mathbf h_i(1)\mathbf q_{1,3}^\perp(1)s_2^{(4)} \nn\\
  &+ \mathbf h_i(1)\mathbf q_{1,2}^\perp(1)s_3^{(4)} + n_i(1),
\end{align}
where $\mathbb E[|\mathbf h_i(1)\mathbf s_1|]\doteq P$. The desired accompanying symbol has power $P^\alpha$ and the remaining two have power $P^0$. Because $\mathbf h_i(1)\mathbf s_1$ is the overlapping of the three included symbols, Rx-$i$ can decode $\mathbf h_i(1)\mathbf s_1$, which contains $(1-\alpha)\log P$ bits of information, at the rate of $(1-\alpha)\log P$ bits/s by considering remaining interference as noise. Afterwards, Rx-$i$ can remove $\mathbf h_i(1)\mathbf s_1$ from its received signal and decode their private symbol at the rate of $\alpha\log P$ bits/s as rest of the interference merges in noise. Using a similar method, in each time slot of Phase~$1$, Rx-$i$ is able to decode $\mathbf h_i(j)\mathbf s_j, j=1,2,3$ as well as their private symbols, i.e., $\mathbf s_i^{(j)}, j=4,5,6$. In order to accomplish the following phases, we repeat the first phase once for new symbol vectors where $\bar{\mathbf s}_1$, $\bar{\mathbf s}_2$ and $\bar{\mathbf s}_3$ replace $\mathbf s_1$, $\mathbf s_2$ and $\mathbf s_3$, respectively, as well as  $s_i^{(j)}, j=7,8,9$ replace $s_i^{(j)}, j=4,5,6$, respectively.

At the end of Phase~$1$, the BS has obtained perfect CSIT, with which it can perfectly reconstruct all the interference terms. Similar to the MAT scheme, the order-$2$ symbols will be reconstructed by using the reconstructed interference. Specifically, $\mathbf h_2(1)\mathbf s_1+\mathbf h_1(2)\mathbf s_2$ is digitalized and coded into $c_{12}$, $\mathbf h_3(1)\mathbf s_2+\mathbf h_1(3)\mathbf s_3$ is digitized and coded into $c_{13}$, as well as $\mathbf h_3(2)\mathbf s_2+\mathbf h_2(3)\mathbf s_3$ is digitized and coded into $c_{23}$. Symmetrically, $\bar{c}_{13}$, $\bar{c}_{23}$ and $\bar{c}_{12}$ are also generated. Because each interference term $\mathbf h_i(j)\mathbf s_j$ is decoded at the rate of $(1-\alpha)\log P$ bits/s of information, according to the rate distortion theorem, each $c_{ij}$ or $\bar{c}_{ij}$ also contains $(1-\alpha)\log P$ bits. In the rest of this section $c_{ij}$ and $\bar{c}_{ij}$ are called {\it order-$2$ common symbols}.

\subsubsection{Phase~$2$}
During this phase, the transmit signal vector, occurring at time slot $t,t=7,8,9$ is given by
\begin{align}
  \mathbf x(t) =& \left[c_{ij}\ \ \bar{c}_{ij}\ \ 0\right]^T + \mathbf q_{2,3}^\perp(t) s_1^{(t+3)}\nn \\
  &+ \mathbf q_{1,3}^\perp(t) s_2^{(t+3)} + \mathbf q_{1,2}^\perp(t) s_3^{(t+3)},
\end{align}
where each order-$2$ common symbol $c_{ij}$ and $\bar{c}_{ij}$ has power $\frac{P}{2}$ in order to satisfy the power constraint. For DoF analysis, since we only care about the power exponent of the received terms, we can safely remove the coefficient $\frac{1}{2}$ found in power of order-2 common symbols.

During these three time slots, each receiver will obtain a combination of $c_{ij}$ and $\bar{c}_i$ with power $P$, a desired private symbols with power $P^\alpha$, and an interference part with power $P^0$. For example, at time slot $7$, Rx-$1$ will receive the following terms
\begin{align}
  y_1(7) &= \underbrace{\mathbf h_1(7)\left[
                                       \begin{array}{c}
                                         c_{12} \\
                                         \bar{c}_{12} \\
                                         0 \\
                                       \end{array}
                                     \right]}_{P} + \underbrace{\mathbf h_1(7)\mathbf q_{2,3}^\perp(7) s_1^{(10)}}_{P^\alpha} \nn\\
                                     &+ \underbrace{\mathbf h_1(7)\mathbf q_{1,3}^\perp(7) s_2^{(10)}}_{P^0} + \underbrace{\mathbf h_1(7)\mathbf q_{1,2}^\perp(7) s_3^{(10)}}_{P^0} + n_1(7).
\end{align}
As the linear combination of $c_{12}$ and $\bar{c}_{12}$ contains $(1-\alpha)\log P$ bits of information, it can be successfully decoded by regarding the rest of the terms as noise. Afterwards, the private symbol $s_1^{(10)}$ can be decoded at the rate of $\alpha\log P$ bits/s by using the successive decoding method. Similarly, at time slot $t,t=7,8,9$, Rx-$i$ will be able to decode the combination of order-$2$ common symbols as well as their private symbols $s_i^{(t+3)}$.

At the end of Phase~$2$, the BS has obtained perfect CSI. Similar to the MAT scheme, it will reconstruct the overheard interference and generate the order-$3$ common symbols, i.e., $\mathbf c_{123}^{(1)}$ is coded from $\mathbf h_3(7)\left[c_{12}\ \ \bar{c}_{12}\ \ 0\right]^T$, $c_{123}^{(2)}$ is coded from $\mathbf h_2(8)\left[c_{13}\ \ \bar{c}_{13}\ \ 0\right]^T$, and $c_{123}^{(3)}$ is coded from $\mathbf h_1(9)\left[c_{23}\ \ \bar{c}_{23}\ \ 0\right]^T$. It is worth mentioning that as each combination of $c_{ij}$ and $\bar{c}_{ij}$ is decoded at the rate of $(1-\alpha)\log P$ bits/s, each $c_{123}^{(i)}$ contains $(1-\alpha)\log P$ bits of information.

\subsubsection{Phase~$3$}
In the two time slots of Phase~$3$, the BS will send two linear combinations of $c_{123}^{(1)}$, $c_{123}^{(2)}$ and $c_{123}^{(3)}$. Specifically, the symbol vector sent at time slot $t$, where $t=10,11$, is given by
\begin{align}
  \mathbf x(t) &= \left[u_t c_{123}^{(1)}+v_t c_{123}^{(2)}+w_t c_{123}^{(3)}\ \ \ 0\ \ \ 0\right]^T  \notag\\
                 & + \mathbf q_{2,3}^\perp(t) s_1^{(t+3)}+ \mathbf q_{1,3}^\perp(t) s_2^{(t+3)} + \mathbf q_{1,2}^\perp(t) s_3^{(t+3)},
\end{align}
where the constants $u_t,v_t,w_t$, for $t=10,11$ are shared with the receivers.

The signal received by Rx-$i$ at time slot $t$ is given by
\begin{align}
  y_i(t) &= \mathbf h_i(t)\!\!\left[\!\!
                   \begin{array}{c}
                     u_t c_{123}^{(1)}+v_t c_{123}^{(2)}+w_t c_{123}^{(3)}\\
                     0 \\
                     0 \\
                   \end{array}
                 \!\!\right] \!\!+ \mathbf h_i(t)\mathbf q_{2,3}^\perp(t)\mathbf s_1^{(t+3)} \notag \\
                 &+ \mathbf h_i(t)\mathbf q_{1,3}^\perp(t) s_2^{(t+3)} + \mathbf h_i(t)\mathbf q_{1,2}^\perp(t) s_3^{(t+3)}+ n_i(t), \notag
\end{align}
where $u_t c_{123}^{(1)}+v_t c_{123}^{(2)}+w_t c_{123}^{(3)}$ can be decoded at the rate of $(1-\alpha)\log P$ bits/s by regarding the rest of the terms as noise. Similar to Phase~$1$ and Phase~$2$, the private symbol vectors can be decoded at the rate of $\alpha\log P$ bits/s by using the successive cancellation approach (SIC) approach.

\subsubsection{Backwards decoding and DoF calculation}
The symbols contained in $\mathbf s_i$ and $\bar{\mathbf s}_i$ can be decoded by Rx-$i$ by using the approach provided in the MAT scheme as all the required order-$2$ and order-$3$ common symbol vectors have already been obtained. Meanwhile, at Rx-$i$, all the new private symbols included in $s_i^{(j)}$ and $\bar{s}_i^{(j)}$, where $j=7,8,\dots,14$, can be decoded in each time slot by using the SIC method.

On one hand, this scheme sends $\mathbf s_1$, $\bar{\mathbf s}_1$, $\mathbf s_2$, $\bar{\mathbf s}_2$, $\mathbf s_3$, $\bar{\mathbf s}_3$ to receivers in $11$ time slots, which sums up to $18(1-\alpha)\log P$ bits of information. Therefore, the total DoF gained for decoding $\mathbf s_1$, $\bar{\mathbf s}_1$, $\mathbf s_2$, $\bar{\mathbf s}_2$, $\mathbf s_3$, $\bar{\mathbf s}_3$ are $\frac{18(1-\alpha)}{11}$. On the other hand, in each of the time slots, $3$ private symbol are sent for all the receivers and each of them contains $\alpha\log P$ bits of information. That is to say, the DoF per time slot for decoding $s_i^{(j)}, j=7,8\dots,14$ is $3\alpha$. Hence, the total DoF achieved by this scheme are
\begin{equation}
  \frac{18(1-\alpha)}{11} + 3\alpha = \frac{18+15\alpha}{11}.\notag
\end{equation}
Given the symmetry of this scheme for Rx-$1$, Rx-$2$, and Rx-$3$, each receiver obtains $\frac{6+5\alpha}{11}$ DoF.
\subsection{Scheme $\mathcal X_2$ (vertices $D_1,D_2,D_3$)}
The Scheme $\mathcal X_2$  describes the procedure  to achieve Vertex $D_3=(1,\alpha,\alpha)$ in the DoF region. At this achievability scheme, occuring  one time slot, the transmit symbol vector  to be sent is
\begin{equation}
  \mathbf x = \left[
                \begin{array}{c}
                  s_1^{(1)} \\
                  0 \\
                  0 \\
                \end{array}
              \right] + \mathbf q_{2,3}^\perp s_1^{(2)} + \mathbf q_{1,3}^\perp s_2^{(1)} + \mathbf q_{1,2}^\perp s_3^{(1)},
\end{equation}
where $s_1^{(1)}$ is a private symbol for Rx-$1$ and $\mathbb E[|\mathbf s_1^{(1)}|^2]\doteq P$. Symbols $s_1^{(2)}$, $s_2^{(1)}$, and $s_3^{(1)}$ are for Rx-$1$, Rx-$2$, and Rx-$3$, respectively, and each of them has power $P^\alpha$. The received signal at Rx-$1$ can be expressed as
\begin{align}
  \!\!\! \!y_1\! = \!\underbrace{h_{1,1} s_1^{(1)}}_{P} \!+\! \underbrace{\mathbf h_1\mathbf q_{2,3}^\perp s_1^{(2)}}_{P^\alpha} \!+\! \underbrace{\mathbf h_1\mathbf q_{1,3}^\perp s_2^{(1)}}_{P^0} \!+\! \underbrace{\mathbf h_1\mathbf q_{1,2}^\perp s_3^{(1)}}_{P^0} \!+\! n_1,
\end{align}
where $h_{1,1}$ denotes the channel entry from the first antenna at the BS to Rx-$1$. Rx-$1$ will decode $s_1^{(1)}$ at rate $(1-\alpha)\log P$ bits/s by considering  the rest of the terms as noise. Then, it can remove $h_{1,1}s_1^{(1)}$ and decode symbol $s_1^{(2)}$ at the rate of $\alpha\log P$ bits/s because the other interference drowns  into the noise level. Therefore, Rx-$1$ gets $(1-\alpha)+\alpha=1$ DoF in total.

On the other side, Rx-$2$ and Rx-$3$ will also be able to decode $h_{i,1} s_1^{(1)}, i=2,3$ at rate $(1-\alpha)\log P$ bits/s. However, they regard it as an undesired symbol vector and remove it from the received signal. Afterwards, they can decode their private symbol at the rate of $\alpha\log P$ bits/s. Hence, each of them obtains $\alpha$ DoF. As a result, the DoF vertex $(1,\alpha,\alpha)$ is achieved.

By changing $s_1^{(1)}$ into the corresponding symbols for Rx-$2$ or Rx-$3$, the vertices $D_1$ and $D_2$ can be achieved, respectively.

\subsection{Extension to $K$-user with $K$-antenna BS scenario}
For the proposed $K$-user setting, two achievability schemes are considered, i.e., Scheme $\mathcal X_3$ and Scheme $\mathcal X_4$. To be more specific, Scheme $\mathcal X_3$  achieves the symmetric DoF vertex, i.e., all the users get the same DoF, while Scheme $\mathcal X_4$  achieves the DoF vertex when one of the users gets {\it one} DoF similar to vertices $D_1,D_2,D_3$ in Figure \ref{fig:region3d_1}.

\subsubsection{Scheme $\mathcal X_3$}
\begin{mytheo}
    In a $K$-user MISO BC, where the BS is equipped with $K$ antennas, each of the receivers is able to achieve
    \begin{equation}
      \frac{1}{1+\frac{1}{2}+\cdots+\frac{1}{K}}+\alpha
    \end{equation}
    DoF per second per Hz.
\end{mytheo}

\textbf{Remark:} This DoF region is consistent with two well known results. Specifically, when $\alpha=0$, the Scheme $\mathcal X_3$ reduces to the MAT scheme, where $\frac{1}{1+\frac{1}{2}+\cdots+\frac{1}{K}}$ DoF are achieved. When $\alpha=1$, the achievable DoF becomes {\it one} per user which can be simply achieved by  using ZF beamforming method.

The MAT scheme is modified,  in order to achieve this DoF vertex. Specifically, Scheme $\mathcal X_3$ can be divided into two parts, the MAT part and the ZF part:
\begin{itemize}
  \item In the MAT part, the MAT scheme is performed but instead of constructing each symbol vector with $\log P$ bits of information, we construct it with $(1-\alpha)\log P$ bits of information. Consequently, each order-$i$ symbol vector used in the MAT scheme, where $i=1,2,\dots,K$ will contain $(1-\alpha)\log P$ bits of information. By using a backwards decoding approach as stated in \cite{Maddah2012}, $\frac{K(1-\alpha)}{1+\frac{1}{2}+\frac{1}{3}+\cdots\frac{1}{K}}$ DoF are achieved.
  \item In the ZF part, since all the symbol vectors in the MAT part  contain $(1-\alpha)\log P$ bits of information, we can send them with power $P$, and save the extra space to send new private symbol vectors at each time slot. Given that  the BS is equipped with $K$ antennas, it can provide $K$ independent $K$-dimensional vectors. Specifically, if $s_i$ is desired by Rx-$i$, the symbol vector in the ZF part is beamformed as
      \begin{equation}
        \sum_{i=1}^K\mathbf q_{1,2,\dots,i-1,i+1\dots,K}^\perp s_i,
      \end{equation}
      where $\mathbf q_{1,2,\dots,i-1,i+1\dots,K}^\perp$ denotes the beamforming vector in the null space of the estimated current channel vectors to all the receivers except Rx-$i$. If $\mathbb E[|s_i|^2]\doteq P^\alpha$, all the overheard interference will be at noise level. Therefore, at each time slot, there are $K$ symbols containing $\alpha\log P$ bits of information for each of the receiver and $K\alpha$ DoF are achieved.
\end{itemize}
As a result, the total DoF are
\begin{equation}
  \frac{K(1-\alpha)}{1+\frac{1}{2}+\frac{1}{3}+\cdots\frac{1}{K}}+K\alpha, \notag
\end{equation}
which are evenly allocated to the $K$ receivers.

\subsubsection{Scheme $\mathcal X_4$}
Scheme $\mathcal X_4$ is a straightforward extension of Scheme $\mathcal X_2$. Specifically, because the BS is equipped with $K$ antennas, it can provide $K$ linear independent $K$-dimensional spaces. Let us denote the beamforming vector for the desired symbol vector for Rx-$i$ as $\mathbf q_{1,2,\dots,i-1,i+1\dots,K}^\perp$, and define new symbols $s_i^{(1)}$ and $s_i^{(2)}$  intended for the same receiver. Then, the beamformed symbol vector can be written as
\begin{equation}
  \mathbf x = \left[s_i^{(1)}\ 0\ \dots\ 0\right]^T
   + \sum_{j=1}^K\mathbf q_{1,2,\dots,j-1,j+1\dots,K}^\perp s_j^{(2)},\notag
\end{equation}
where $\mathbb E\left[|s_i^{(1)}|^2\right]\doteq P$ and $\mathbb E\left[|s_i^{(2)}|^2\right]\doteq P^\alpha, \forall i$. For Rx-$k$, $k=1,2,\dots,K$, the received signals are
\begin{align}
   y_k &= \underbrace{h_{k,1}s_i^{(1)}}_{P} + \underbrace{\mathbf h_k\mathbf q_{1,2,\dots,k-1,k+1\dots,K}^\perp s_k^{(2)}}_{P^\alpha}\nn \\
   &+ \underbrace{\mathbf h_k\sum_{j=1, j\neq k}^K\mathbf q_{1,2,\dots,j-1,j+1\dots,K}^\perp s_k^{(2)}}_{P^0}+n_k.
\end{align}
Rx-$k$ can first decode $h_{k,1}s_i^{(1)}$ at rate $(1-\alpha)\log P$ bits/s by regarding the rest of the terms as noise. Then it can remove it from $y_k$ and decode its own private symbol at rate $\alpha\log P$ bits/s because the remaining interference is at noise level.

Taking into account that  $s_i^{(1)}$ is intended for Rx-$i$,  the achieved DoF at Rx-$i$ are $(1-\alpha)+\alpha=1$. At all the other receivers, the achieved DoF are $\alpha$.

\section{Assymetric Case ($N=2$,  $K=3$)}  \label{sec:case2}
In this section, we present the achievable DoF region and achievability schemes for a $3$-user MISO BC, where the BS is equipped with 2 antennas.

\subsection{DoF Region}
\begin{figure}[t]
  \centering
  \includegraphics[width=0.9\textwidth]{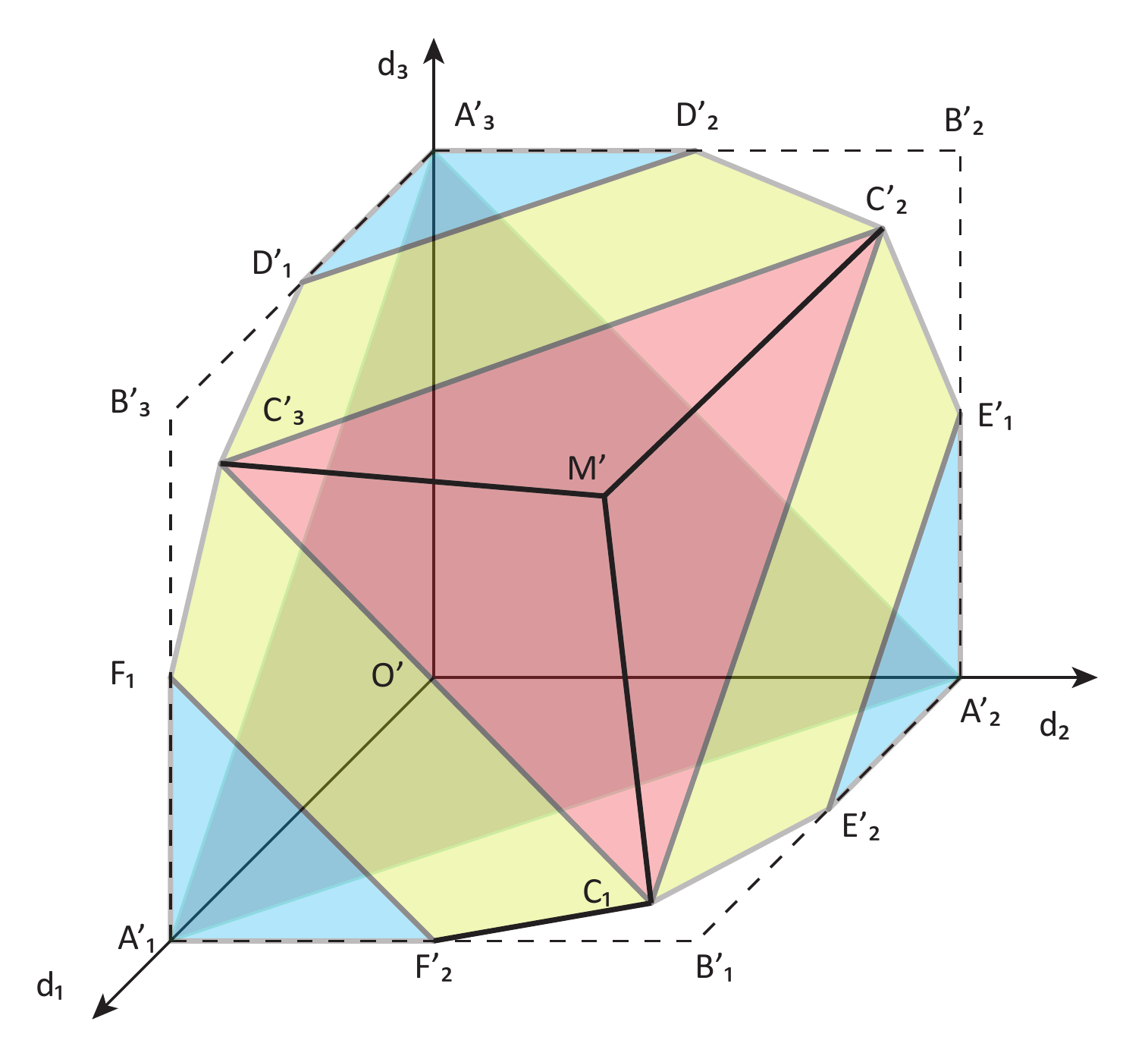}\\
  \vspace{-2mm}
  \caption{The DoF polygon for the asymmetric case ($M=2$,  $K=3$).}\label{fig:region3d_2}
  \vspace{-2mm}
\end{figure}

\begin{mytheo}
The DoF region, achieved by the $3$-user MISO BC with perfect delayed  as well as imperfect current CSIT when the BS is equipped with $2$ antennas, can be formed by the following inequalities, i.e.,
\begin{align}
  &d_1\leq 1,\ d_2\leq 1,\ d_3\leq 1, \notag\\
  &2d_1+d_2+d_3\leq 2+\alpha, \notag\\
  &d_1+2d_2+d_3\leq 2+\alpha, \notag\\
  &d_1+d_2+2d_3\leq 2+\alpha, \notag\\
  &2d_1+3(1+\alpha)d_2+3(1+\alpha)d_3\leq (2+\alpha)(3+\alpha), \notag\\
  &3(1+\alpha)d_1+2d_2+3(1+\alpha)d_3\leq (2+\alpha)(3+\alpha), \notag\\
  &3(1+\alpha)d_1+3(1+\alpha)d_2+2d_3\leq (2+\alpha)(3+\alpha), \notag\\
  &d_1+d_2+d_3\leq 2. \notag
\end{align}
All the important vertices are listed and the formed polygon is shown in Fig.~\ref{fig:region3d_2}. Specifically,  the vertices are given by
\begin{align}
  C'_1&=\left(\frac{2+\alpha}{3},\frac{2+\alpha}{3},0\right), C'_2=\left(0,\frac{2+\alpha}{3},\frac{2+\alpha}{3}\right), \notag\\
C'_3&=\left(\frac{2+\alpha}{3},0,\frac{2+\alpha}{3}\right), \notag \\
  M'&=\left(\frac{3+\alpha}{6},\frac{3+\alpha}{6},\frac{3+\alpha}{6}\right). \notag
\end{align}
\end{mytheo}

\textbf{Remark:} The DoF polygon represented by the  triangle $A'_1A'_2A'_3$ can be obtained by the  scheme, where {\it one} DoF can be achieved totally. All the remaining vertices except Vertex $O'$ achieve $2$ total DoF when $\alpha=1$. The total DoF achieved at vertex $M'$ is $\frac{3+\alpha}{2}$, which is consistent to two well known results, i.e., when $\alpha=1$, $2$ DoF are achieved and when $\alpha=0$, $\frac{3}{2}$ DoF are obtained, as proved in \cite{Maddah2012}.

\subsection{Scheme $\mathcal X_5$ for vertex $M'$}
The total scheme is divided into three phases where the first two phases consist of $3$ time slots, while the last phase includes $2$ time slots. The two symbol vectors, sent to each of the receivers, are $\mathbf s_i=[s_i^{(1)}\ s_i^{(2)}]^T, \bar{\mathbf s}_i=[\bar{s}_i^{(1)}\ \bar{s}_i^{(2)}]^T$ for Rx-$i$, $i=1,2,3$.

\subsubsection{Phase~$1$}
In the first phase, the BS sends signal vectors to two of the three receivers at each time slot, i.e., the symbol vector sent in the first time slot can be written as
\begin{equation}
  \mathbf x(1) = \left[\mathbf q_2(1)\ \ \mathbf q^\perp_2(1)\right]\mathbf s_1 + \left[\mathbf q_1(1)\ \ \mathbf q^\perp_1(1)\right]\mathbf s_2, \notag
\end{equation}
which contains no symbol vector intended for Rx-$3$. Symbol $\mathbf q_i(t)$ and $\mathbf q_i^\perp(t)$ denote the vectors in the span and the null space of $\mathbf h_i(t)$, respectively. The power allocation of each of the symbol vectors are $\mathbb E[|s_1^{(1)}|^2]\doteq\mathbb E[|s_2^{(1)}|^2]\doteq P^{1-\alpha}$ and $\mathbb E[|s_1^{(2)}|^2]\doteq\mathbb E[|s_2^{(2)}|^2]\doteq P$. Symmetrically, the symbol vectors, sent in the following two time slots, are given by
\begin{align}
  \mathbf x(2) &= \left[\mathbf q_3(2)\ \ \mathbf q^\perp_3(2)\right]\bar{\mathbf s}_1 +\left[\mathbf Q_1(2)\ \ \mathbf q^\perp_1(2)\right]\mathbf s_3, \notag \\
  \mathbf x(3) &= \left[\mathbf q_3(3)\ \ \mathbf q^\perp_3(3)\right]\bar{\mathbf s}_2
  + \left[\mathbf Q_2(3)\ \ \mathbf Q^\perp_2(3)\right]\bar{\mathbf s}_3. \notag
\end{align}

At the receiver side, the received symbols and the corresponding power allocations in the first time slot are given by
\begin{align}
   y_1(1) &= \underbrace{\mathbf h_1(1)\mathbf q_2(1) s_1^{(1)}}_{P} + \underbrace{\mathbf h_1(1)\mathbf q^\perp_2(1) s_1^{(2)}}_{P^{1-\alpha}} + \underbrace{\eta_1}_{P^{1-\alpha}} + \underbrace{n_1(1)}_{P^0},\notag \\
   y_2(1) &= \underbrace{\mathbf h_2(1)\mathbf q_1(1) s_2^{(1)}}_{P} + \underbrace{\mathbf h_2(1)\mathbf q^\perp_1(1) s_2^{(2)}}_{P^{1-\alpha}} + \underbrace{\eta_2}_{P^{1-\alpha}} + \underbrace{n_2(1)}_{P^0},\notag
\end{align}
where $\eta_1$ and $\eta_2$ describe the interference at Rx-$1$ and Rx-$2$, respectively. Note that Rx-$3$ regards all its received signal as interference. The total power of the interference terms at Rx-$1$ is given by
\begin{equation}
   \eta_1 = \underbrace{\mathbf h_1(1)\mathbf q_1(1) s_2^{(1)}}_{P^{1-\alpha}} + \underbrace{\mathbf h_1(1)\mathbf q^\perp_1(1) s_2^{(2)}}_{P^{1-\alpha}},
\end{equation}
by which we have $\mathbb E[|\eta_1|^2]\doteq P^{1-\alpha}$. Similarly, $\eta_2$ has the same power.  At the end of each time slot, the BS will obtain the perfect CSIT of this time slot, which enables it to reconstruct the interference terms, i.e., $\eta_1$ and $\eta_2$. In order to make room for privates symbol that will be sent in later phases, each of the reconstructed interference will be digitalized into $(1-\alpha)\log P$ bits to make the digitalization error $\tilde{\eta}_i$ submerged in white noise. After the digitalization process, the digitalized version of $\eta_i$ is included in the new common symbol $c_i$ that will be sent in the next phase, where $\mathbb E[|c_i|^2]\doteq P$.

The symbols received in the other two time slots in this phase have similar form with the received signal  as in the first time slot. In general, the BS reconstructs $6$ interference terms and each one will be digitalized into $(1-\alpha)\log P$ bits. Note that the other common symbols $c_i$ and $\bar{c}_i$, where $i=1,2,3$, contain interference overheard by Rx-$i$.

\subsubsection{Phase~$2$}
The second phase also consists of $3$ time slots and the symbol vectors sent at each time slot are symmetric. Specifically, in the $4$-th time slot, the transmit symbol vectors  are given by
\begin{equation}
  \mathbf x(4) = \left[
                      \begin{array}{c}
                        c_1 \\
                        c_2 \\
                      \end{array}
                    \right]
                    +\mathbf q_2^\perp(4) s_1^{(3)}+\mathbf q_1^\perp(4) s_2^{(3)},
\end{equation}
where $s_1^{(3)}$ and $s_2^{(3)}$ are new private symbols for Rx-$1$ and Rx-$2$, respectively, and $\mathbb E[|s_1^{(3)}|^2]\doteq\mathbb E[|s_2^{(3)}|^2]\doteq P^\alpha$.

Note that  Rx-$i$, $i=1,2$,  can decode $h_{1,1}(4) c_2 + h_{1,1}(4) c_1$ at the rate of $(1-\alpha)\log P$ bits/s and remove it from its received signal. Then it can decode its desired symbol at $\alpha\log P$ bits/s. After obtaining perfect CSIT at the end of the $4$-th time slot, the BS will generate new symbols by reconstructing interference terms observed by Rx-$3$ as $d_1$ and $d_2$. Note that the symbol $d_i$ or $\bar{d}_i$ is the digitalized symbol for the  interference term containing $c_i$ and $\bar{c}_i$, respectively.

In the remaining two time slots, the BS will reconstruct the interference terms overheard at Rx-$2$, Rx-$3$, respectively. Consequently, digitalized symbols $\bar{d}_1$, $\bar{d}_2$, $d_3$, and $\bar{d}_3$, representing the interference terms, are generated.

\subsubsection{Phase~$3$}
The symbol vector transmitted at the $t$-th time slot, where $t=7,8$, is given by
\begin{align}
  \mathbf x(t) &\!=\! u_t\left[\!
                     \begin{array}{c}
                       \bar{d}_2+\bar{d}_3 \\
                       0 \\
                     \end{array}
                   \!\right]
                   \!+\! v_t\left[\!
                     \begin{array}{c}
                       \bar{d}_1+ d_3 \\
                       0 \\
                     \end{array}
                   \!\right]
                   \!+\! w_t \left[\!
                     \begin{array}{c}
                       d_1+d_2 \\
                       0 \\
                     \end{array}
                   \!\right]
                   \!+\! \mathbf c_t, \notag
\end{align}
where $u_i,v_i,w_i,i=7,8$ are shared  all  receivers and the common symbol $\mathbf c_t,t=7,8$, where $\mathbb E[\|\mathbf c_t\|^2]\doteq P^\alpha$, contains two private symbol vectors similar to the second phase. To keep the scheme symmetric among all the receivers, private symbol vectors contained in $\mathbf c_t$ are evenly transmitted to  the three receivers. The most simple way to evenly allocate the symbols is to repeat the whole scheme three times and in each time, $\mathbf c_t,t=7,8$ is destined  to one receiver. By regarding the common symbol vector $\mathbf c_t$ as noise, the  receivers will be able to decode the sum of all the other terms at the rate of $(1-\alpha)\log P$ bits/s. After that, the private symbol can be decoded at the rate of $\alpha\log P$ bits/s.

\subsubsection{Backwards decoding and DoF calculation}
Similarly, the scheme can be divided into the MAT part and the ZF part. For the MAT part, the exact way to decode $\mathbf s_i$ and $\bar{\mathbf s}_i$, where $i=1,2,3$ refers to the MAT scheme. Note that each $\mathbf s_i$ or $\bar{\mathbf s}_i$ contains two symbols with $\alpha\log P$ bits and $\log P$ bits each. Therefore, the achievable DoF  per time slot in the MAT part are $\frac{6(1-\alpha+1)}{8}=\frac{6-3\alpha}{4}$.

Meanwhile, for the ZF part, at each time slot, two new symbols are sent and each of them is decoded at the rate of $\alpha\log P$ bits/s. Altogether, during the last two phases, $10$ new private symbols are sent, each of which contains $\alpha\log P$ bits of information. Specifically in Phase~$2$, the . Thus, the total DoF achieved per time slots are $\frac{10\alpha}{8}=\frac{5\alpha}{4}$.

Adding the DoF achieved in the MAT  and the ZF parts, the total achievable DoF  are
\begin{equation}
  \frac{6-3\alpha}{4} + \frac{10\alpha}{8} = \frac{3+\alpha}{2}.
\end{equation}
Note that the total DoF are symmetrically allocated to Rx-$1$, Rx-$2$, and Rx-$3$.

\section{Conclusion}    \label{sec:conclusion}
This paper proposed the DoF regions and achievability schemes in MISO time-correlated BC by taking into account for  imperfect current  as well as perfect delayed CSIT. In total, $5$ achievability schemes are provided for $2$ DoF regions, i.e., a $K$-user and $K$-antenna BS (symmetric) case as well as a $3$-user and $2$-antenna (asymmetric) case.
\begin{itemize}
  \item \textbf{$K$-User and $K$-Antenna BS:} How to achieve the DoF outer bound with $K$-user and $K$-antenna BS with imperfect current  and perfect delayed CSIT was an open problem. We proposed the Scheme $\mathcal X_3$ and achieved $\frac{K(1-\alpha)}{1+\frac{1}{2}+\frac{1}{3}+\cdots\frac{1}{K}}+K\alpha$ DoF, which is equal to the known published DoF outer bound. By  varying $\alpha$, Scheme $\mathcal X_3$ compromised between the MAT scheme and the ZF beamforming method     . That is to say, the current CSIT is always beneficial. Moreover, we proposed the Scheme $\mathcal X_4$, which provided one receiver {\it one} DoF and all the other receivers $\alpha$ DoF. Note that $K$ DoF can be achieved in total when $\alpha=1$.
  \item \textbf{$3$-User and $2$-Antenna BS:} Firstly, a $3$-dimensional DoF polygon consisting of $12$ surfaces was illustrated. Afterwards, Scheme $\mathcal X_5$ was proposed, which achieved $\frac{3+\alpha}{2}$ DoF in total, and therefore, the scheme is consistent with two well known results, i.e., when $\alpha=0$, a total of $\frac{3}{2}$ DoF are achieved as obtained by the MAT scheme, while when $\alpha=1$, $2$ DoF can be achieved as obtained by using the ZF method in case of only perfect current CSIT.
\end{itemize}

\section*{Acknowledgment}
This research was supported by HIATUS and HARP projects, as well as a Marie Curie Intra European Fellowship within the 7th European Community Framework Programme for Research of the European Commission under grant agreements no. [265578], HIATUS and  no. [318489], HARP, as well as  no. [330806], IAWICOM.

\bibliographystyle{IEEEtran}

\end{document}